\documentstyle[aps,prl,epsf]{revtex}
\def\limite{\mathop\rightarrow}

\begin{document}

\title{
Warm and cold Denaturation in the
Phase Diagram of a Protein Lattice Model.
}

\author{Olivier Collet}

\address{Laboratoire de Chimie Th\'eorique, UMR CNRS 7565,
Facult\'{e} des Sciences, \\
Universit\'{e} Henri Poincar\'{e}-Nancy 1,
54506 Vandoeuvre-l\`{e}s-Nancy, France}
\address{
\centering{
\medskip\em
{}~\\
\begin{minipage}{14cm}
Studying the properties of the solvent around
proteins, we propose a much more sophisticated model of solvation than
temperature-independent pairwise interactions between monomers,
as is used commonly in lattice representations. 
We applied our model of solvation to a 16-monomer chain 
constrained on a two-dimensional lattice.
We compute a phase diagram function of the temperature and 
a solvent parameter which is related to the pH of the solution.
It exhibits a native state in which the
chain coalesces into a unique compact conformation as well as a denatured state.
Under certain solvation conditions, both warm and cold denaturations 
occur between the native and the denatured states.  
A good agreement is found with the data obtained from 
calorimetric experiments, thereby validating the proposed model. 
{}~\\
{}~
\end{minipage}
}
}

\maketitle
\newpage

Under physiological condition, a protein adopts a  unique 
three dimensional structure which is totally encoded in its
amino acid sequence \cite{Anfinsen1961}.
Because, its biological function is closely related to 
this native structure,
an important challenge in molecular modeling is to predict the
native structure of a protein, given its amino acid sequence.
It is now commonly assumed that the solvent plays a very important
role in the folding process of proteins towards their native
state \cite{Kauzmann1959,Privalov1989}. Its contribution to the 
conformational free energy should, therefore, be
introduced in the computation of the energy in simulations  of
biological molecules \cite{Brooks1988,Premilat1997}.

Many proteins are in their native form between $10^\circ {\rm C}$
and $40^\circ {\rm C}$ \cite{Privalov1992}.
At higher temperatures, a reversible, warm denaturation occurs.
For these proteins, a cold denaturation also occurs at low temperatures.
For example, Privalov \cite{Privalov1992} showed experimentally that, 
under medium pH condition, the evolution of the heat capacity of myoglobin 
as a function of the temperature exhibits two peaks
that are the signatures of two modifications in the order of the chain:
it only folds at medium temperatures, and both a cold and a warm denaturation 
occurs. 

A widely used class of models for studying theoretically 
protein folding is the lattice representation \cite{Shakhnovich1990a,Lau1989}.
Up until now, solvent effects were taken into account in these models 
using a parameter independent of the temperature, introduced 
in the intrachain calculations and
favoring compact conformations. Dinner {\it et al.} \cite{Dinner1994}
used the random energy  model (REM) \cite{Derrida1981}
to compute phase diagrams of protein-like chains. With this solvent model,
they found three different regions: a native state, a globular state and
a random coil state.
Their phase diagrams, however,  do not exhibit any sign
characteristic of a cold denaturation.
Several more sophisticated solvent models have been already proposed in 
the literature 
\cite{Pratt1977,Pohorille1990,Chandler1993,Garde1996,Hummer1998a,Gomez1999,Sorenson1999},
but they have not yet been able to simulate a cold denaturation in a protein model.

Two theoretical analyses of disordered phases at low temperature 
have been proposed recently.
Hansen {\it et al.} \cite{Hansen1998}
used a zipper model equivalent to the 1-D Ising model, and obtained warm and cold transitions.
De Los Rios and Caldarelli \cite{DeLosRios1999} proposed
a Potts model that allows the observation of two peaks in the heat capacity
profiles of short chains but the warm unfolding transition
disappears as the length of the chain increases. Moreover, 
both models are homopolymer models, and, as such, do not
exhibit regions in which the chain folds into a unique,
compact, native, structure.

In this letter, we present a simple solvation model for REM, based on 
a study of the thermodynamic and the structural properties
of the aqueous environment of a protein. 
We find a phase diagram which exhibits a native and a denatured regions with
both warm and cold denaturation transitions.
Our results are qualitatively in good agreements with the experimental findings 
of Privalov \cite{Privalov1992}.
We apply our solvation model to a 16-monomers chain constrained on
a two-dimensional lattice for which 802075 different self avoiding walk 
conformations non equivalent by symmetry can be easily enumerated.
We introduce a parameter $B_s$ in the model (more details are given below)
which is related to the quality and the pH of the
solvent and acts as a control parameter. 
At small $B_{s}$, the protein has a strong tendency to fold,
while at large $B_s$ solvation becomes more effective and
 the protein tends to unfold.

\begin{figure}[htbp]
\epsfxsize=1.5in
\epsfysize=1.6in
\centerline{\hskip-1.6in \epsffile{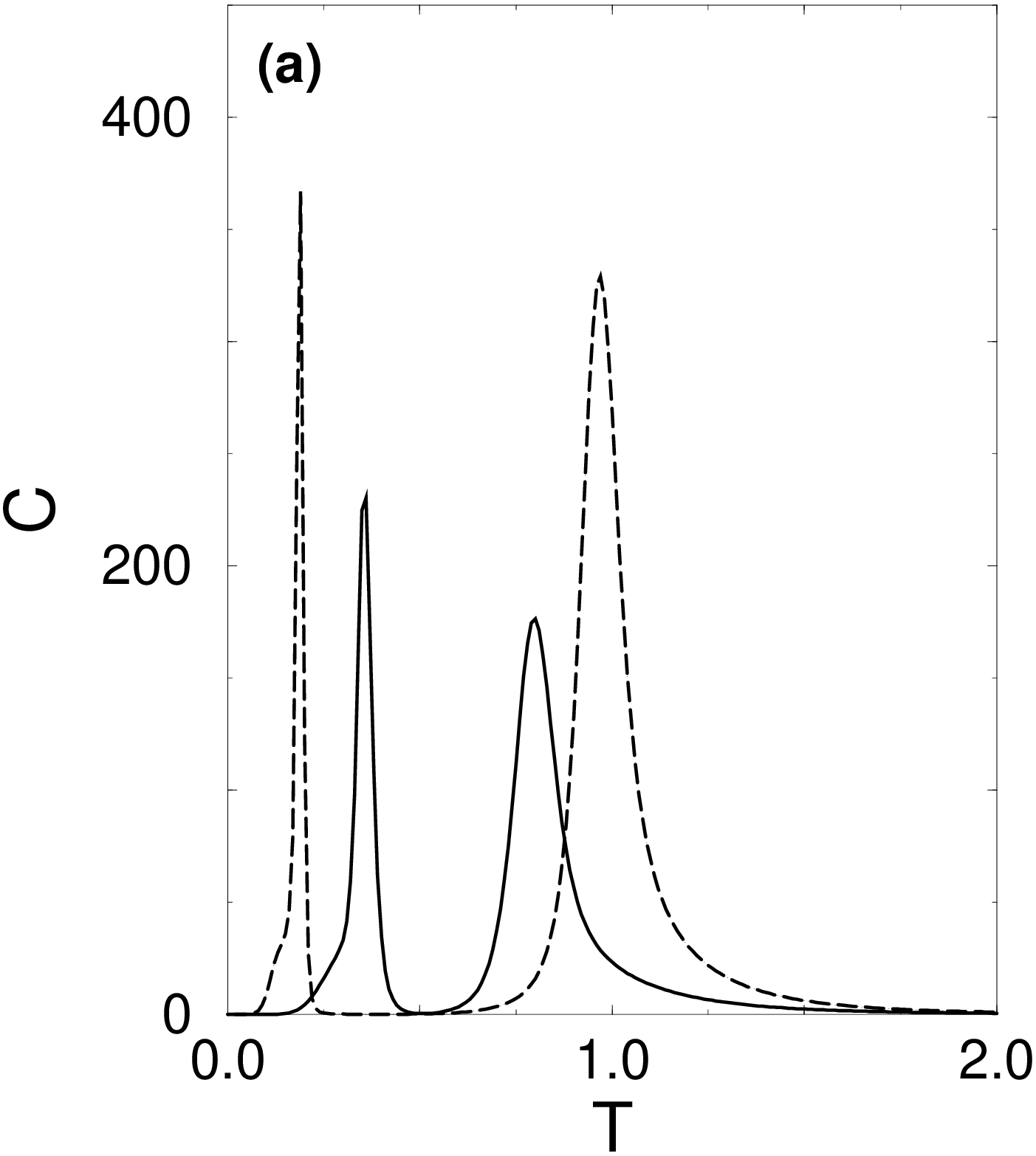}}
\epsfxsize=1.5in
\epsfysize=1.6in
\centerline{\hskip-1.6in \epsffile{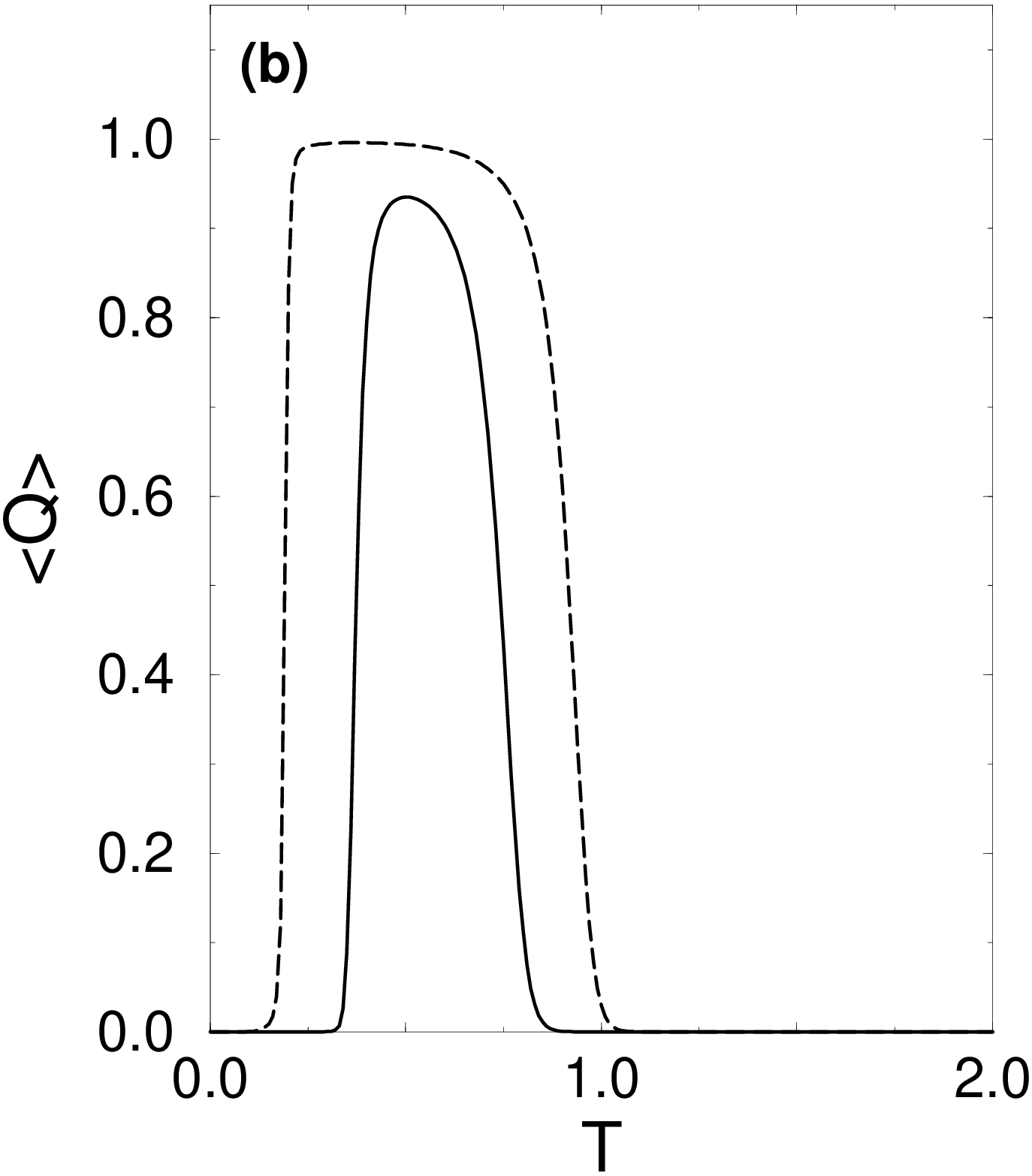}}
\vskip-3.2in
\epsfxsize=2.1in
\epsfysize=1.6in
\centerline{\hskip+1.6in \epsffile{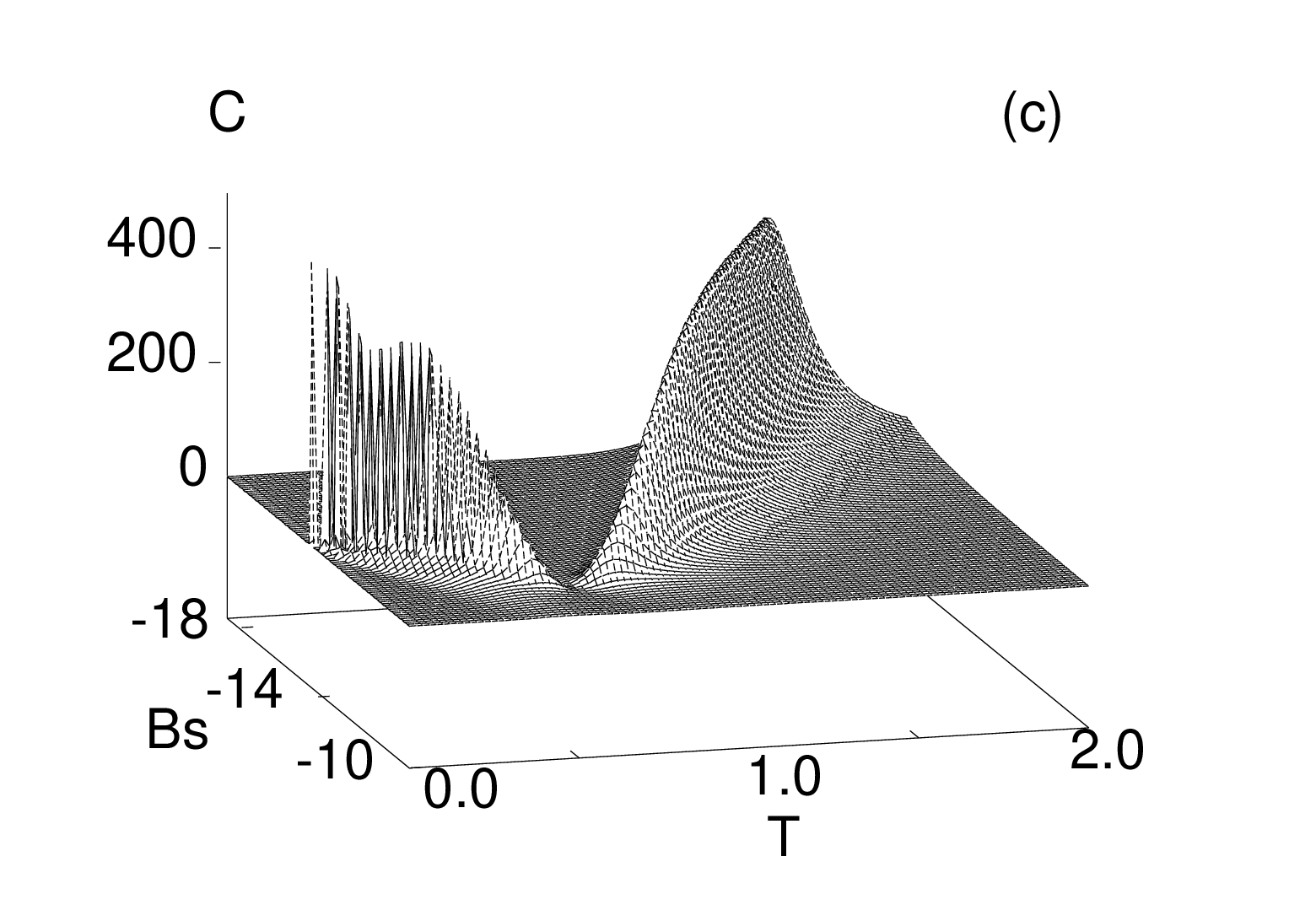}}
\epsfxsize=2.1in
\epsfysize=1.6in
\centerline{\hskip+1.6in \epsffile{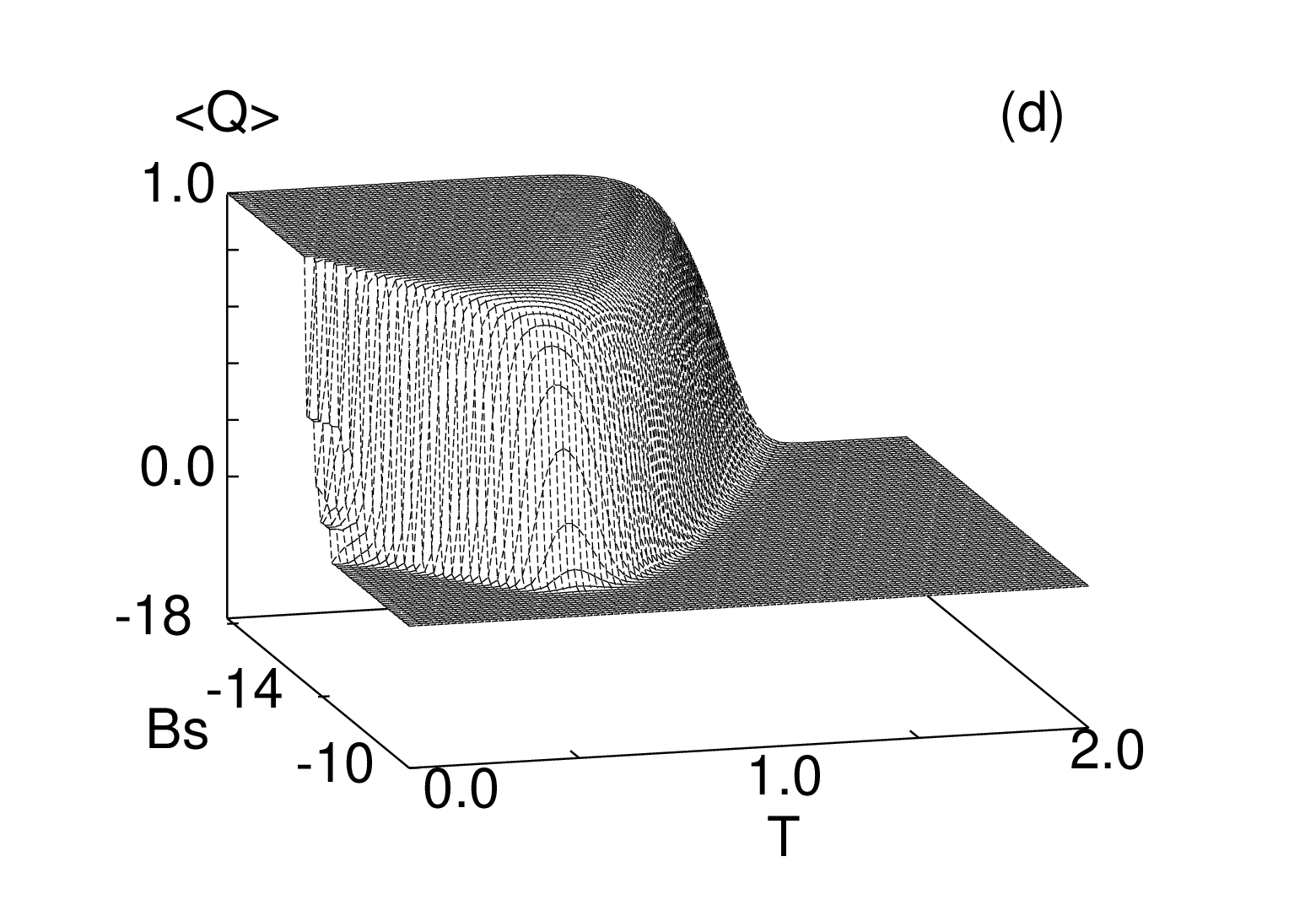}}
\epsfxsize=1.4in
\epsfysize=2.0in
\centerline{\hskip-1.6in \epsffile{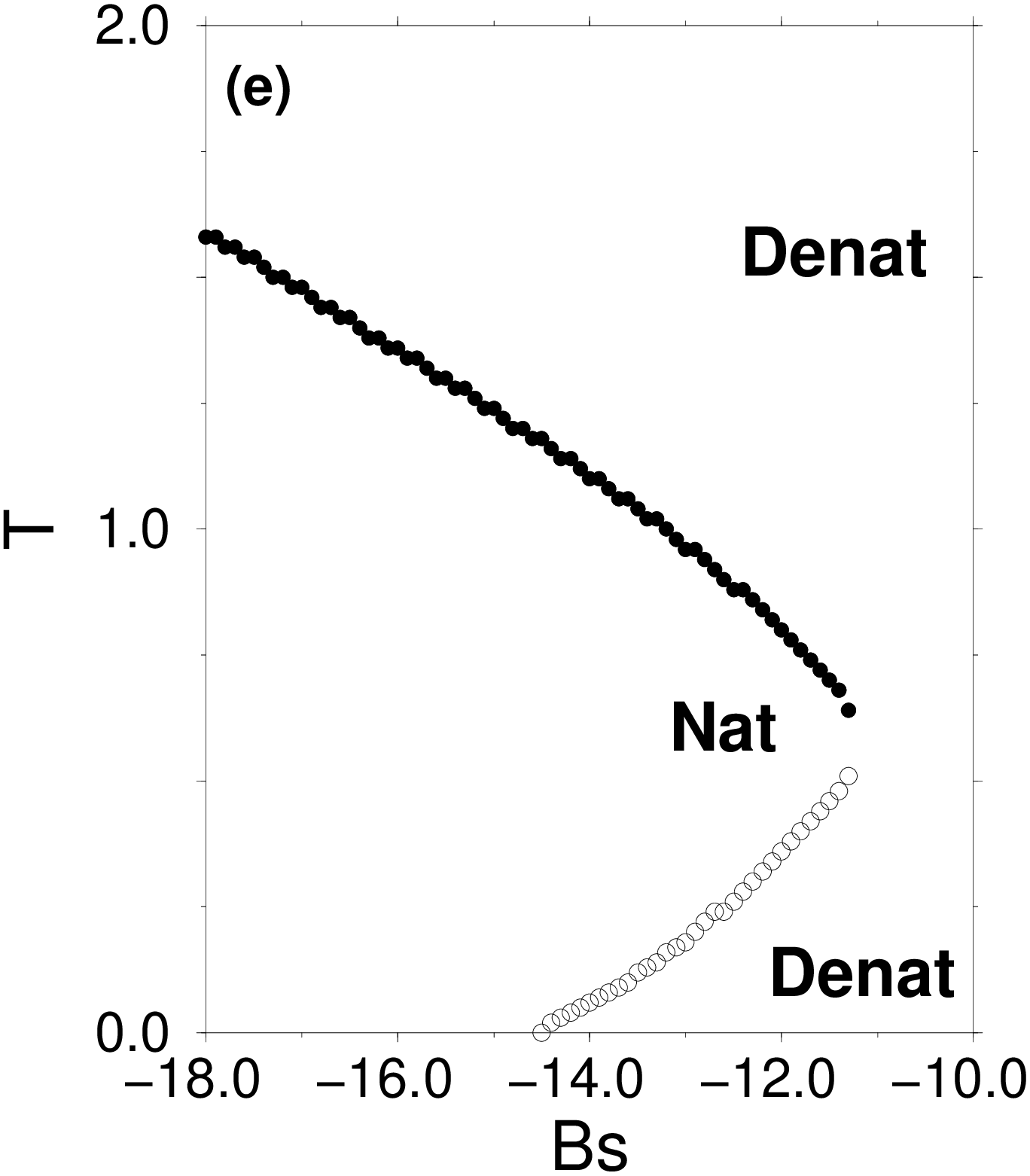}}
\vskip-2.0in
\epsfxsize=1.4in
\epsfysize=2.0in
\centerline{\hskip+1.6in \epsffile{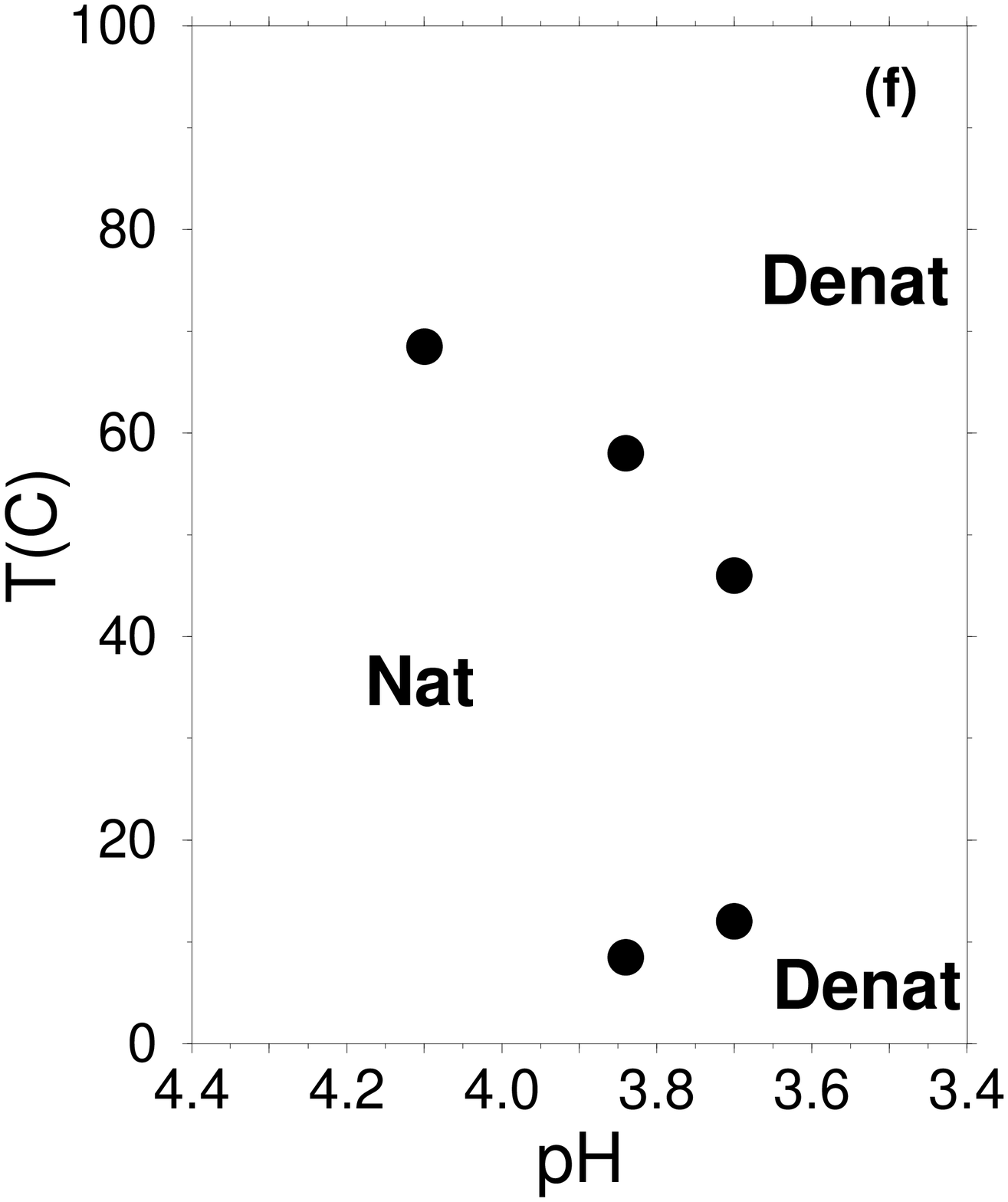}}
\epsfxsize=0.3in
\epsfysize=0.3in
\vskip-0.92in
\centerline{\hskip-2.0in \epsffile{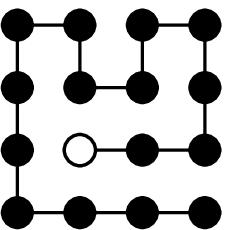}}
\vskip0.8in
\caption{
(a) Heat capacity curve and
(b) order parameter
as functions of the temperature for $B_{s}$
equals -12.0 (solid lines) and -13.0 (dotted lines).
(c) Heat capacity curves and
(d) order parameter
as a function of the temperature and the solvent parameter.
In (e) and (f) Nat is for the native region and Denat is for the denatured region.
(e) Phase diagram of the model calculated with the present work
for a chain of $N=16$ monomers.
The filled and empty symbols indicate warm and cold denaturation
transition respectively. 
The native structure is shown. The empty circle is for the 
first monomer
(f) Phase diagram of myoglobin obtained experimentally
by Privalov. No transition is found for the pH 
below 3.5 where  myoglobin is always denatured.
}
\label{phase}
\end{figure}

The main results of our investigation are summarized in fig.\ref{phase}. 
Fig.\ref{phase}(a), (b), (c) and (d) shows the behavior of the 
specific heat $C$ and of the order parameter $\langle Q \rangle$.
Both quantities are plotted versus temperature (fig.\ref{phase}(a) 
and (b)) and versus temperature and solvation parameter 
(fig.\ref{phase}(c) and (d)).
The two peaks on the heat capacity curves shown on fig.\ref{phase}(a) 
for $B_s = -12.0$ and $B_{s} = -13.0$ are the 
signatures of the modification of the order of the chain.
A warm denaturation occurs at the temperature $T_w$
corresponding to the higher temperature heat capacity peak.
The temperature of the other peak, when occurs, is the temperature of cold 
denaturation $T_c$. 
The sharp decay of the order parameter 
indicates that the transitions are first order type.
The fact that, $\langle Q \rangle = 1 $ between $T_c$ and $T_w$, 
shows that the chain is in a unique native compact structure.
$\langle Q \rangle \rightarrow 0$ at temperature below than $T_c$ 
or above than $T_w$ then a huge number of extended conformations
of the chain are relevant.
One must note the particularity of the cold denaturation transition~:
the chain is in a disordered state at low temperature and becomes
ordered in a unique structure as the temperature increases.

No peaks occurs, for values of $B_{s}>-11.2$ (fig.\ref{phase}(c))
and the chain is always denatured (fig.\ref{phase}(d)).
The peak corresponding to the cold denaturation do not appear for
$B_{s}<-14.4 $ and, only the warm denaturation still occurs.
The curves $T_c(B_{s})$ and $T_w(B_{s})$ are reported
on the ($B_s$, $T$) diagram of the system (fig.\ref{phase}(e)).
One sees that for $(B_s, T)=(-11.2, 0.58)$ the two curves
intersect. At this particular point a naturation at
constant temperature occurs by modifying the solvent 
condition without increasing the heat capacity.
This transition seems occur without latent heat.
Phase diagram for myoglobin obtained from result of Privalov
are shown on fig.\ref{phase}(f).
The good agreement between the calorimetric experiments
and our theoretical results strongly supports our 
model of solvation that we introduce now.

\begin{figure}[htb]
\epsfxsize=1.5in
\epsfysize=2.0in
\centerline{\hskip-1.6in \epsffile{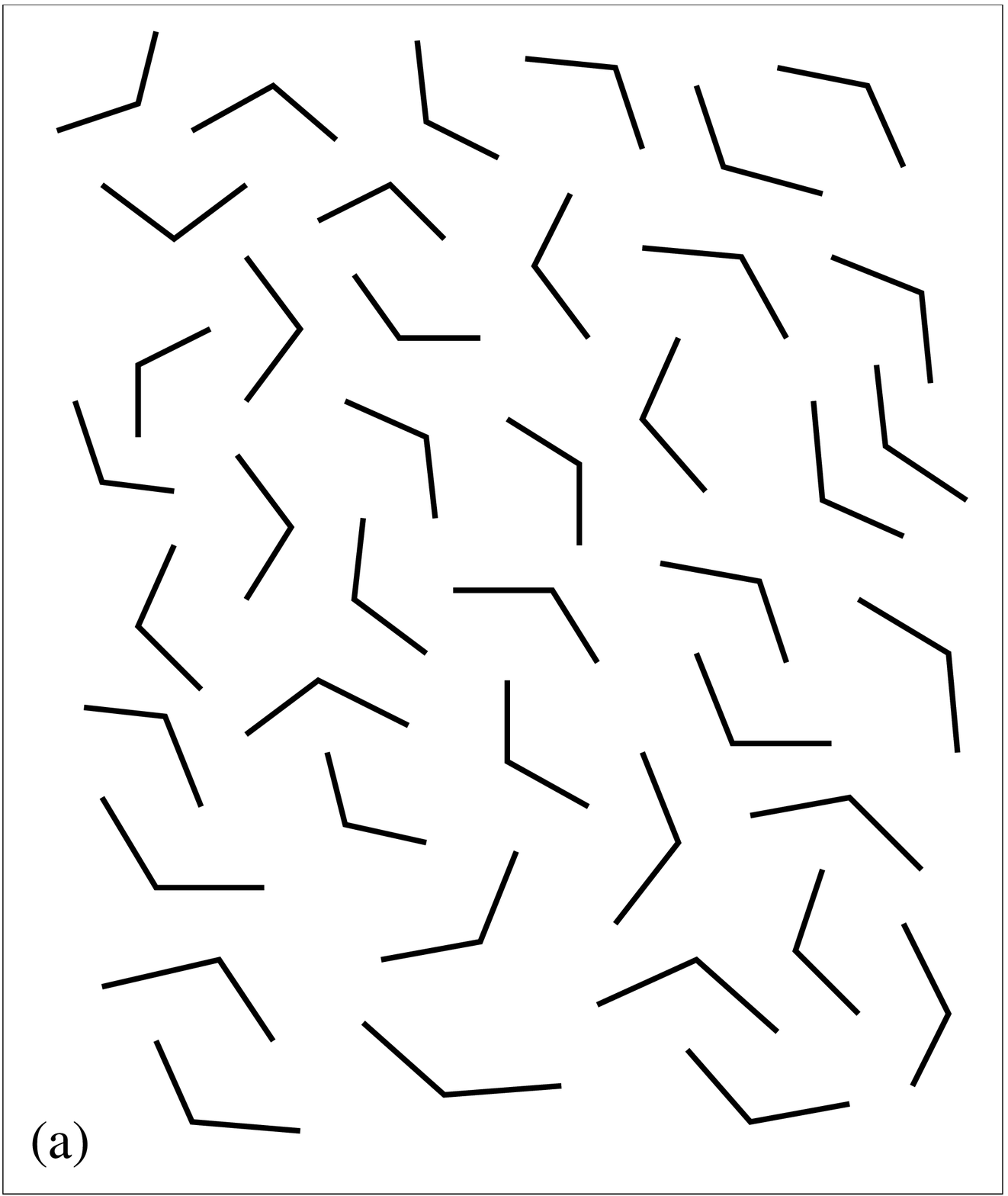}}
\vskip-2.0in
\epsfxsize=1.5in
\epsfysize=2.0in
\centerline{\hskip1.6in \epsffile{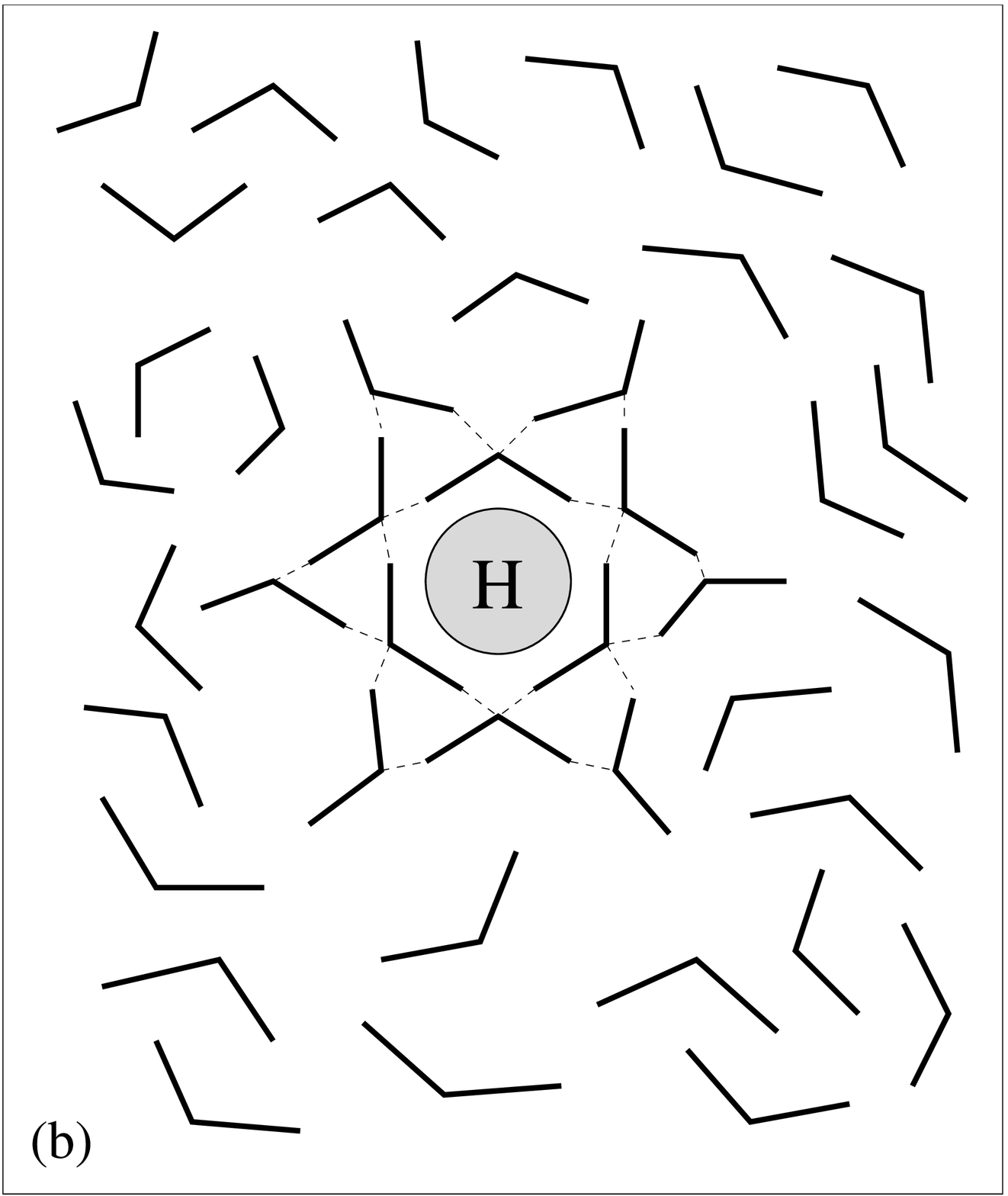}}
\caption{
Water molecules are represented by angles.
(a) In the absence of hydrophobic compounds, the system of
water molecules evolves very freely and has a large entropy.
(b) An hydrophobic solute H interacts only weakly
with water molecules. Then, the solvent molecules are
organized around the compound to form
a very rigid water cage (a clatracle).
Hence, the entropy of the system (b) is lower than
that of the system (a) and therefore the more hydrophobic compounds
are hidden from the solvent in the interior of the
protein to increase the conformational entropy of
solvation.
}
\label{water}
\end{figure}

The mechanism driving proteins towards compact structures
is strongly coupled to the hydrophobic effect \cite{Kauzmann1959}. 
The latter arises from a loss of solvent entropy
when a hydrophobic monomer is transfered from the interior of the
protein to the vicinity of the aqueous solvent (fig.\ref{water}). 
In addition, intrachain interactions play the ultimate role for finding the
unique, native structure among all the compact structures
involving internal organization.
Let us note $E_{\rm intr}^{(m)}$
the intrachain energy of the chain in the conformation $m$ and 
$E_{\rm solv}^{(mm')}$ the solvation energy resulting of
the interactions between the solvent molecules 
in the conformation $m'$ and between solvent molecules and 
the chain units.
The partition function of the system in thermal equilibrium reads~:
\begin{equation}
Z(T) = \sum_{m \in \Omega} \, \ \sum_{m' \in \Omega'(m)} \exp
\left(-\frac{E_{\rm intr}^{(m)} + E_{\rm solv}^{(mm')}}{T}\right)
\end{equation}
where $\Omega$ is the conformational space of the chain and 
one must note that the conformational space of the solvent $\Omega'(m)$ 
depends on the chain conformation $m$. 
The free energy of solvation  is written~:
\begin{equation}
F_{\rm solv}^{(m)}(T) = - T \ln \sum_{m' \in \Omega'(m)} \exp
\left(-\frac{E_{\rm solv}^{(mm')}}{T}\right)
\end{equation}
Then, it comes~: 
\begin{equation}
Z(T) = \sum_{m \in \Omega} \exp \left(- \frac{F_{\rm tot}^{(m)}(T)}{T}\right)
\end{equation}
where the total free energy of the conformation $m$ is simply~:
$ F_{\rm tot}^{(m)}(T) = E_{\rm intr}^{(m)} + F_{\rm solv}^{(m)}(T)$

The chain is represented by a string 
of $N$ beads constrainted on a square lattice. 
The intrachain energy is given by the classical expression~:
\begin{equation}
\label{Eintr}
E_{\rm intr}^{(m)} = \sum_{i=1}^N \sum_{j > i}^N B_{ij} \ \Delta_{ij}^{(m)} =
\frac{1}{2} \sum_{i=1}^N \sum_{j \ne i}^N B_{ij} \ \Delta_{ij}^{(m)}
\end{equation}
where $\Delta_{ij}^{(m)}$ equals 1 if the monomer $i$ and $j$ are
first neighbors on the lattice and the symmetric couplings $B_{ij}=B_{ji}$ are chosen
at random in a gaussian distribution with a standard deviation equals 1.
In contrast with previous works \cite{Shakhnovich1990a} the mean of the gaussian law
is taken equals to 0 in order to generate 
attractive and repulsive interactions between the monomers. 

Each empty site of the lattice is filled up with a group of solvent molecules
which interacts with its four first neighbor sites on which are located  
either monomers of the chain or other groups of solvent molecules.
Then, the free energy of solvation  is written as the sum 
of interactions involving at least one solvent sites (solvent-monomer interactions 
and solvent-solvent interactions)~:
\begin{equation}
\label{fsolv}
F_{\rm solv}^{(m)}(T) = \sum_{i=1}^N n_i^{(m)} f_{i}(T) + n_{s}^{(m)}
f_{s}(T)
\end{equation}
where $n_i^{(m)}$ is the number of solvent sites first neighbors of
the $i^{th}$ monomer on the lattice and $n_{s}^{(m)}$ is the total number of
first neighbors interactions between any pairs of solvent sites.
Both quantities depend on the conformation of the chain. 
$f_{i}(T)$ is the contribution to the free energy of the interaction between a
group of solvent molecules and the monomer $i$ 
and $f_{s}(T)$ is the free energy of interaction between two solvent sites.
They are given by~:
\begin{equation}
\label{fi}
f_{i}(T) = -T \ln \sum_{j=1}^{N_s} \exp(-B_{i}^{(j)}/T)
\ \ \ {\rm and} \ \ \
f_{s}(T) = -T \ln \sum_{j'=1}^{N_s} \exp(-B_{s}^{(j')}/T)
\end{equation}
where $B_{i}^{(j)}$ is the energy of the bond between the 
$i^{th}$ monomer and a group of solvent molecules in the conformational state $j$ and
the state $j'$ of the bond between two solvent sites has for energy
$B_{s}^{(j')}$.
The two summations run over the conformational states of the
system composed of the solvent molecules of one group.
We assume that, the number of states $N_s$ which can adopt the system of 
solvent molecules is the same in both cases, but,  the solvation energy
density functions of $B_{i}$ and $B_{s}$ are different (fig.\ref{water}).
We give below the arguments for choosing these functions.

We assume that the density energy function of the interaction between the solvent 
and a monomer is well represented by a gaussian law.
The non-degenerated energy minimum of the $N_s$ states 
corresponds to the clatracle conformation (fig.\ref{water}b) and at sufficiently 
low temperature, only this state is relevant.
The mean of the gaussian law is taken as the solvation energy
reference and is chosen equals to 0. The $N_s$ states, i.e. the 
$N_s$ values of $B_i$,  are chosen at 
random in the gaussian law with a standard deviation equals 2. 
The value of $B_{i}^{\rm min} = \min_j {B_{i}^{(j)} }$\cite{bmin} is specific 
for each amino acid and is the energy of the solute surrounded
by the clatracle conformation.  It gives an insight of the 
hydrophobicity of the monomer $i$.

On the other hand, one assumes that in pure solvent,
the water molecules can evolve very freely and then,
the $N_s$ conformational states of a bond between two sites of 
solvent molecules have the same energy (fig.\ref{water}a), let says $B_{s}$.
$f_s(T)$ can then be rewritten~: $f_{s}(T) = B_{s} - T \ln N_s$
where the $B_s$ parameter is related to the pH of solution as follow~: 
when pH of the solution equals 7, the water molecules form a large number
of hydrogen bonds, then $B_s$ is very negative.
As pH of the solvent is lowered from 7, more and more hydrogen bonds
between water molecules are broken and then the solvation energy $B_s$ 
increases but is still negative. 
Then, the larger $B_s$, the lower the pH of the solution.

Let us now show that, the total free energy of conformation can be
rewritten as the sum of effective pairwise interactions between monomers.
The total number of lattice links, $n_{\rm tot}$ can be written as the sum of 
number of covalent bonds, intrachain contacts, monomer-solvent sites 
interactions and solvent-solvent sites contacts~:
$n_{\rm tot} = (N-1) + \sum_{i=1}^N \sum_{j>i}^N \Delta_{ij}^{(m)} +
	\sum_{i=1}^N n_i^{(m)} + n_{s}^{(m)} $
Let us note, $\alpha_i$ is the number of first neighbors
the $i^{th}$ monomer except the next and previous ones 
($\alpha_1 = \alpha_N = 3$ and $\alpha_i = 2$ for $2 \le i \le N-1$).
One has~:
$ n_i^{(m)} = \alpha_i - \sum_{j \ne i} \Delta_{ij}^{(m)} $
Then, following eqs. \ref{Eintr}, \ref{fsolv}
and the free energy of conformation  reads~: 
\begin{equation}
\label{Ftot2}
  F_{\rm tot}^{(m)}(B_{s},T) = F_{\rm tot}^{(ext)}(B_{s},T) +
    \sum_{i=1}^N \sum_{j \ne i}^N \left( \frac{1}{2} B_{ij}
    - f_i(T) + \frac{1}{2} f_s(B_s,T) \right) \Delta_{ij}^{(m)}
\end{equation}
where $F_{\rm tot}^{(ext)}(B_{s},T)= (n_{\rm tot}-3N-1) f_s(T) + \sum_i \alpha_i f_i(T)$ 
is the free energy of the extended chains. For these chains, the intrachain energy
vanishes and the total free energy is only the contribution of the free energy
of solvation.
As, $\sum_i \sum_{j \ne i} f_i(T) \ \Delta_{ij}^{(m)} = \sum_i \sum_{j \ne i} 
\frac{1}{2} (f_i(T)+f_j(T)) \ \Delta_{ij}^{(m)}$, eq.\ref{Ftot2} can be rewritten~: 
\begin{equation}
  F_{\rm tot}^{(m)}(B_{s},T) = F_{\rm tot}^{(ext)}(B_{s},T) + 
        \sum_{i=1}^N \sum_{j>i}^N B_{ij}^{\rm eff} (B_s, T)
  \ \Delta_{ij}^{(m)}
\end{equation}
where $B_{ij}^{\rm eff}(B_s,T)$ are the effective couplings which now, 
dependent on the temperature~: 
\begin{equation}
B_{ij}^{\rm eff} (B_{s},T)=B_{ij}-f_{i}(T)-f_{j}(T)+f_{s}(B_{s},T)
\end{equation}
One must note that the free energy of conformation is rewritten as
the sum of temperature-dependent pairwise interactions between monomers.

This temperature dependence of the couplings, that we study now, induces the 
cold denaturation phenomenon.
In eq.13, the solvation contribution of the $i^{\rm th}$ monomer to the
couplings $B_{ij}^{\rm eff} (B_{s},T)$ is 
$\delta_i(T) = -f_i(T) + f_s(B_s,T)/2$.
The derivatives of $\delta_i(T)$ reads, $\partial \delta_i(T) / \partial T
= s_i -s_s/2$ where $s_s = \ln N_s$ is the entropy of the group of neat solvent
and $s_i(T)$ is the entropy of the solvent around the solute $i$.
One has $s_i(0)=0$ and 
$s_i(T) \displaystyle \limite_{T \rightarrow \infty} \ln N_s$.
It comes~: $\partial \delta_i(0)/ \partial T = -\ln(N/2)$ and
$\partial \delta_i(T) / \partial T \displaystyle \limite_{T \rightarrow 
\infty} \ln (N/2)$.
Then, there is a value of $T_i$ for which $\partial \delta_i(T_i) / \partial T = 0$
and for which $\delta_i(T)$ has a maximum value.
And, as $\delta_i(0)=-B^{\rm min}_i+B_s/2$, for some values of $B_s$, the contribution
$\delta_i(T)$ is positive for small and large value of $T$, 
inducing on averaged, repulsive interactions of monomer $i$
and, in contrast, $\delta_i(T)$ has negative values for medium values of the 
temperature and compact structures are then favored.

For each value of ($B_{s}, T$) an order parameter,
is computed and define the state of the chain~:
\begin{equation}
\langle Q (B_{s}, T) \rangle =
\frac{1}{N_{\rm max}}  \sum_{m_1 \in \Omega} \sum_{m_2 \in \Omega}
N_{\rm com}^{(m_1m_2)} \ P_{eq}^{(m_1)} \ P_{eq}^{(m_2)}
\end{equation}
where the two summations are taken over all the chain conformations and
$N^{(m_1 m_2)}_{\rm com}$ is the number of  common contacts between the
conformations $m_1$ and $m_2$, $N_{\rm max}$ is the number
of contacts of the more maximally compact conformation (here 
$N_{\rm max} = 9$) and $P_{eq}^{(m)}(B_{s}, T)
=\exp[-F_{\rm tot}^{(m)}(B_{s}, T) / T]/Z(B_{s}, T))$
is the equilibrium probability of occurence of the conformation $m$. 
Hence, when only one structure $m$ is relevant, one has $P_{\rm eq}^{(m)}=1$
and then $\langle Q (B_{s}, T) \rangle = N^{(m)}/N_{\rm max}$, with $N^{(m)}$
is the number of contacts of the conformation $m$.
If the conformation $m$ is more maximally compact, one has 
$\langle Q (B_{s}, T) \rangle = 1$. On the other hand, if a large number of structure have a non-null probability of occurence, one has~:
$\langle Q (B_{s}, T) \rangle \rightarrow 0$.

The $B_{ij}$ values are those of the sequence R given in \cite{Cieplak1997}
after centering the mean of these values on 0.
In order to fully solvate the more extended chain, one supposes that
at least $2N+2$ groups of solvent molecules surround the polymer.
The thermal averaged of the quantities of interest are computed
for each couple of values of $B_s$ and $T$ such as $B_s$ vary 
from -5.00 to -9.00 with a step of 0.05 and $T$ varying from 0.00 to 2.00 
with a step of 0.02.  One generates sets of $N_s=10000$ values of 
$B_{i}^{(j)}$ for each amino acid to compute the partial free energies 
of solvation $f_i(T)$, following eq.\ref{fi}.
Phase diagrams computed with $N_s = 2000, 5000, 10000, 20000$ remain
qualitatively unchanged. The boundary between the two regions is only
shifted along the $B_{s}$ axis. 

In conclusion, 
we have presented a new model of solvation based on thermodynamic and structural
properties of the aqueous solvent around protein. We showed, that the
interactions involving in this solvation model can be written as 
temperature-dependent pairwise interaction between monomers
including, the temperature and a solvent parameter which can be related
to the pH of the solution.
The model has been applied to a two-dimensional chain and we obtained a phase
diagram that exhibits  warm and cold denaturations and which  is
qualitatively in good agreement with experimental results.
It must also be noted that our solvation model is easily applicable to longer 
chains on a three-dimensional lattice.

{\bf Acknowledgments} to Dragi Karevski
for helpful discussions to Enrico Carlon and Christophe Chipot for critical 
reading of the manuscript and to Centre Charles Hermite, 
Vandoeuvre-l\`es-Nancy, France for allocation of computer time.


\newpage

\newpage
 
\begin{thebibliography}{10}

\bibitem{Anfinsen1961}
C.~B. Anfinsen, E. Haber, M. Sela, and W. F.H., Proc. Natl. Acad. Sci. {\bf
  47},  1309  (1961).

\bibitem{Kauzmann1959}
W. Kauzmann, Adv. Protein Chem. {\bf 14},  1  (1959).

\bibitem{Privalov1989}
P.~L. Privalov and G.~I. Makhatadze, J. Mol. Biol. {\bf 205},  737  (1989).

\bibitem{Brooks1988}
C.~L. Brooks, {III}, M. Karplus, and B.~M. Pettitt, {\em Proteins} (John Wiley
  \& Sons, New York, 1988).

\bibitem{Premilat1997}
S. Premilat and O. Collet, Europhysics Letters {\bf 39},  575  (1997).

\bibitem{Privalov1992}
P. Privalov,  in {\em Protein Folding}, edited by T.~E. Creighton (W. H.
  Freeman and Company, New York, 1992), pp.\ 83--126.

\bibitem{Shakhnovich1990a}
E.~I. Shakhnovich and A.~M. Gutin, Nature {\bf 346},  773  (1990).

\bibitem{Lau1989}
K.~F. Lau and K.~A. Dill, Macromolecules {\bf 22},  3986  (1989).

\bibitem{Dinner1994}
A. Dinner, A. \v{S}ali, M. Karplus, and E. Shakhnovich, J. Chem. Phys. {\bf
  101},  1444  (1994).

\bibitem{Derrida1981}
B. Derrida, Phys. Rev. B {\bf 24},  2613  (1981).

\bibitem{Pratt1977}
L. Pratt and D. Chandler, J. Chem. Phys. {\bf 67},  3683  (1997).

\bibitem{Pohorille1990}
A. Pohorille and L. Pratt, J. Am. Che. Soc. {\bf 112},  5066  (1990).

\bibitem{Chandler1993}
D. Chandler, Phys. Rev. E {\bf 48},  2898  (1993).

\bibitem{Garde1996}
S. Garde {\it et~al.}, Phys. Rev. Lett. {\bf 77},  4966  (1996).

\bibitem{Hummer1998a}
G. Hummer {\it et~al.}, J. Phys. Chem. B {\bf 102},  10469  (1998).

\bibitem{Gomez1999}
M.~A. Gomez, L.~R. Pratt, G. Hummer, and S. Garde, J. Phys. Chem. B {\bf 103},
  3520  (1999).

\bibitem{Sorenson1999}
J.~M. Sorenson {\it et~al.}, J. Phys. Chem. B {\bf 103},  5413  (1999).

\bibitem{Hansen1998}
A. Hansen, M.~H. Jensen, K. Sneppen, and G. Zocchi, European Physical Journal B
  {\bf 6},  157  (1998).

\bibitem{DeLosRios1999}
P. De~Los~Rios and G. Caldarelli, cond-mat/9903394  .

\bibitem{bmin}
$B^{\rm min}_i$ for $i=1$ to $16$ : -7.73 ; -7.97 ; -7.54 ; -8.16 ; -7.43 ;
  -7.79 ; -7.36 ; -8.20 ; -6.98 ; -7.58 ; -7.70 ; -8.18 ; -7.48 ; -8.14 ; -7.84
  ; -7.61  .

\bibitem{Cieplak1997}
M. Cieplak and J.~R. Banavar, Folding \& Design {\bf 2},  235  (1997).

\end{thebibliography}
\end{document}